\documentstyle[12pt]{article}

\begin{document}

\title{\bf
Anisotropic simple-cubic Ising lattice:
extended phenomenological renormalization-group treatment
}

\author{
M.~A.~Yurishchev
\thanks{Electronic address: yur@icp.ac.ru}
}
\date{
Institute of Problems of Chemical Physics
of the Russian Academy of Sciences,
142432 Chernogolovka,
Moscow Region,
Russia
}

\maketitle

\begin{abstract}
Using transfer-matrix extended phenomenological renormalization-group
methods [6] the improved estimates for the critical temperature of
spin-1/2 Ising model on a simple-cubic lattice with partly anisotropic
coupling strengths ${\vec J}=(J',J',J)$ are obtained.
Universality of both fundamental critical exponents $y_t$ and $y_h$
is confirmed.
We show also that the critical finite-size scaling amplitude ratios
$A_{\chi^{(4)}}A_\kappa/A_\chi^2$, $A_{\kappa^{''}}/A_\chi$, and
$A_{\kappa^{(4)}}/A_{\chi^{(4)}}$ are independent of the lattice
anisotropy parameter $\Delta=J'/J$.
\end{abstract}

\medskip



\newpage
\section{Introduction}
\label{sec:Intro}
The phenomenological renormalization-group (RG) method in which the
transfer-matrix technique and finite-size scaling (FSS) ideas
are combined is a powerful tool for investigation of critical
properties in different two-dimensional systems \cite{Nigh82,B83}.
Unfortunately, its application in three and more dimensions is
sharply retarded due to huge sizes of transfer matrices which
arise in approximations of $d$-dimensional lattices by
$L^{d-1}\times\infty$ subsystems.

Indeed, even in the simplest case of systems with only two
states of a site (a spin-${1\over2}$ Ising model) the
order of transfer matrix in three dimensions ($d=3$) increases
according to the law $2^{L^2}$ (instead of the essentially more
sparing law $2^L$ in two dimensions).
Hence, for the cluster $3\times3\times\infty$ it is needed to
solve the eigenproblem of the transfer matrix $512\times512$,
for the $4\times4\times\infty$ subsystem ---
$65\,536\times65\,536$, and for the $5\times5\times\infty$
cluster it is required to find the eigenvalues and eigenvectors
of dense matrices with huge sizes of
$33\,554\,432$ by $33\,554\,432$.

One can now solve the full eigenproblem for the transfer matrices of
Ising parallelepipeds $L\times L\times\infty$ with the side length
$L\le4$.
Our aim in this paper is to use such solutions with the largest
effect and extract as much as possible accurate information about
physical properties of the bulk system.

The ordinary phenomenological RG is based on the FSS equations for
correlation lengths \cite{Nigh82,B83}.
However, it is known \cite{DSS81,PFG99,PV00} that the phenomenological
RG can be built up using other quantities with a power divergence
at the phase-transition point.
It is remarkable that the such modified renormalizations can provide
more precise results by the same sizes of subsystems \cite{Yu00}.

In this article we give improved estimates for the
critical temperature of the anisotropic three-dimensional
(3D) Ising model by use of the extended phenomenological RG schemes
found in \cite{Yu00}.
Achieved estimates are taken to calculate the values of different
invariants of the 3D Ising universality class.
Some of them are also given in the paper.


\section{Basic equations}
\label{sec:BE}
Start from the ordinary FSS equations \cite{Nigh82,B83} for the
inverse correlation length $\kappa_L(t,h)$ and singular part of the
dimensionless free-energy density $f^s_L(t,h)$, but we write
them out for the derivatives with respect to the reduced temperature
$t=(T-T_c)/T_c$ and external field $h$:
\begin{equation}
   \label{eq:kappa^mn}
   \kappa_L^{(m,n)}(t, h)
   = b^{my_t+ny_h-1}\kappa_{L/b}^{(m,n)}(t', h')
\end{equation}
and
\begin{equation}
   \label{eq:f^smn}
   f_L^{s\,(m,n)}(t, h)
   = b^{my_t+ny_h-d}f_{L/b}^{s\,(m,n)}(t', h').
\end{equation}
Here
$\kappa_L^{(m,n)}(t,h)=\partial^{m+n}\kappa_L/\partial t^m\partial h^n$
and the same for $f_L^{s\,(m,n)}$; $y_t$ and $y_h$ are,
respectively, the thermal and magnetic critical exponents of the system;
$b=L/L'$ is the rescaling factor.
Deriving Eqs.~(\ref{eq:kappa^mn}) and (\ref{eq:f^smn})
we used a linearized form of RG equations $t'\simeq b^{y_t}t$ and
$h'\simeq b^{y_h}h$.

In the traditional phenomenological RG theory \cite{Nigh82,B83},
Eq.~(\ref{eq:kappa^mn}) with $m=n=0$ is considered as an RG mapping
$(t,h)\to (t',h')$ for a cluster pair $(L,L')$.
By this, the critical temperature $T_c$ is estimated from the
equation
\begin{equation}
   \label{eq:kappaTc}
   L\kappa_L(T_c)=L'\kappa_{L'}(T_c).
\end{equation}

Phenomenological renormalization $(t,h)\to (t',h')$ can be
also realized by using any of the relations (\ref{eq:kappa^mn})
and (\ref{eq:f^smn}) or their combination.
It has been shown by the author \cite{Yu00} that some of such
extended renormalizations lead to more rapid convergence
in $L$ than the standard phenomenological RG transformation.
In particular, test examples on the fully isotropic systems
\cite{Yu00} exhibited that the
relations
\begin{equation}
   \label{eq:kappa2}
   \frac{\kappa_L^{''}}{L^{d-1}\chi_L}\Big|_{T_c}=
   \frac{\kappa_{L'}^{''}}{(L')^{d-1}\chi_{L'}}\Big|_{T_c}
\end{equation}
and
\begin{equation}
   \label{eq:chi4}
   \frac{\chi_L^{(4)}}{L^d\chi_L^2}\Big|_{T_c}=
   \frac{\chi_{L'}^{(4)}}{(L')^d\chi_{L'}^2}\Big|_{T_c}
\end{equation}
locate $T_c$ 
more accurately in comparison with the ordinary RG
equation (\ref{eq:kappaTc}).
In the relations (\ref{eq:kappa2}) and (\ref{eq:chi4}), the derivative
$\kappa_L^{''}=\partial^2\kappa_L/\partial h^2$, the zero-field
susceptibility $\chi_L=f_L^{s\,(0,2)}$, and the nonlinear
susceptibility $\chi_L^{(4)}=f_L^{s\,(0,4)}$ can be evaluated by
standard formulas via the eigenvalues and eigenvectors of transfer
matrices (see, e.~g., \cite{Yu94,Yu97}).

To get the thermal critical exponent $y_t$ we applied two approaches.
Firstly, we used again the standard finite-size expression
\begin{equation}
   \label{eq:yt1}
   y_t=\frac{\ln[L\dot\kappa_L/(L'\dot\kappa_{L'})]}{\ln(L/L')}
\end{equation}
which follows from Eq.~(\ref{eq:kappa^mn}) by $m=1$, $n=0$;
$\dot\kappa_L=\partial\kappa_L/\partial t$.
Secondly, we took the formula
\begin{equation}
   \label{eq:yt2}
   y_t=\frac
   {\kappa_{L'}\dot\kappa_L-\kappa_L\dot\kappa_{L'}}
   {(\kappa_L\kappa_{L'}\dot\kappa_L\dot\kappa_{L'})^{1/2}\ln(L/L')}.
\end{equation}
This expression  is a direct sequence of the well-known Roomany-Wyld
approximant to the Callan-Symanzik $\beta$-function \cite{B83}.

To calculate the magnetic critical exponent $y_h$ we also used two
ways:
\begin{equation}
   \label{eq:yh1}
   y_h={d\over2}+\frac{\ln(\chi_L/\chi_{L'})}{2\ln(L/L')}
\end{equation}
and
\begin{equation}
   \label{eq:yh2}
   y_h={1\over2}+\frac{\ln(\kappa_L^{''}/\kappa_{L'}^{''})}{2\ln(L/L')}
\end{equation}
[these finite-size relations follow from Eqs.~(\ref{eq:kappa^mn}) and
(\ref{eq:f^smn})].

In addition, we calculated the universal ratios of critical FSS
amplitudes.
Such a kind of the ratios can be identified from the Privman-Fisher
functional expressions \cite{PF84} which for partly
anisotropic systems read \cite{Yu97}
\begin{equation}
   \label{eq:kappa}
   \kappa_L(t, h)
   = L^{-1}G_0{\cal K}(C_1tL^{y_t}, C_2hL^{y_h})
\end{equation}
and
\begin{equation}
   \label{eq:fs}
   f_L^{s}(t, h)
   = L^{-d}G_0{\cal F}(C_1tL^{y_t}, C_2hL^{y_h}).
\end{equation}
Scaling functions ${\cal K}(x_1, x_2)$ and ${\cal F}(x_1, x_2)$ are
the same within the limits of a given universality class but they may
depend on the boundary conditions and the subsystem shape (a cube,
infinitely long parallelepipeds, etc.).
All nonuniversality including the lattice anisotropy is absorbed in
the geometry prefactor $G_0$ and metric coefficients $C_1$ and $C_2$.
The critical amplitude ratios from which the parameters $G_0$,
$C_1$, and $C_2$ drop out should be universal.
In particular, the amplitude combinations
\begin{equation}
   \label{eq:Q}
   U=\frac{A_{\chi^{(4)}}A_\kappa}{A_\chi^2}=
   \frac{\kappa_L\chi_L^{(4)}}{L^{d-1}\chi_L^2}
\end{equation}
(Binder-like ratio for the spatially anisotropic systems),
\begin{equation}
   \label{eq:Y1}
   Y_1=\frac{A_{\kappa^{''}}}{A_\chi}=
   \frac{\kappa_L^{''}}{L^{d-1}\chi_L},
\end{equation}
and
\begin{equation}
   \label{eq:Y2}
   Y_2=\frac{A_{\kappa^{(4)}}}{A_{\chi^{(4)}}}=
   \frac{\kappa_L^{(4)}}{L^{d-1}\chi_L^{(4)}}
\end{equation}
are expected to be not depend on the lattice anisotropy parameter
$\Delta=J'/J$.


\section{Results and discussion}
\label{sec:RD}
In the present paper we carried out calculations for the
subsystems $L\times L\times\infty$ with $L=3$ and 4.
To avoid undesirable surface effects the periodic boundary conditions
have been imposed in both transverse directions of parallelepipeds
$L\times L\times\infty$.
Thus, the transfer matrices for which the eigenproblems was solved
were dense matrices of sizes up to $65\,536\times 65\,536$.
To solve the eigenproblem we took into account the internal and
lattice symmetries of subsystems and used the block-diagonalization
method (see, e.\ g., \cite{Yu94}).
Calculations were performed on an 800 MHz Pentium III PC running
the FreeBSD operating system.

Note that the matrix quasidiagonalization procedure is rather
tedious in realization.
Another possible way for treating the full eigenproblem of such
large-scale matrices is to use supercomputers (say,
Ref.~\cite{DFGM01} where the direct diagonalizations have been carried
out for the Hamiltonian matrices of sizes $2^{15}$ by $2^{15}$).


\subsection{Critical temperature}
\label{sec:CT}
The critical temperature estimates coming from solutions of the
transcendental equations (\ref{eq:kappa2}) and (\ref{eq:chi4})
are collected in Table~\ref{tab:Tc}.

In the purely isotropic case $(J'=J)$ there are high precision
numerical estimates for the critical point of the 3D Ising model.
The most precise values for it have been obtained by
Monte Carlo simulations \cite{HPV99,BST99};
$K_c=0.221\,654\,59(10)$, i.\ e. $k_BT_c/J=1/K_c=4.511\,5240(21)$.

Inspecting Table~\ref{tab:Tc} one can see that the estimates
for $J'=J$ which follow from Eq.~(\ref{eq:kappa2}) and
(\ref{eq:chi4}) are the lower and upper bounds respectively.
By this, their mean value has the accuracy of $0.01\%$.
Note also that our mean estimate is better than the
value $k_BT_c/J=4.533\,71$ obtained in Ref.~\cite{N91} (see
also~\cite{BGHP92}) for the fully isotropic lattice by using the
ordinary phenomenological renormalization of the bars with $L=4$
and 5.

Discuss now the anisotropic case.
Here there is well-known exact asymptotic formula for the critical
temperature \cite{WGFF67}
\begin{equation}
   \label{eq:TcQ1D}
   (k_BT_c/J)_{asym}=2/[\ln(J/2J') - \ln\ln(J/2J') + O(1)]
\end{equation}
as $J'/J\to0$.
It is a direct consequence of the molecular-field approximation
in which the linear Ising chain is taken as a cluster.

Unfortunately, simple formula (\ref{eq:TcQ1D}) yields considerable
errors in the region $10^{-3}\le J'/J\le1$.
Its modifications in spirit of Ref.~\cite{GL81},
$k_BT_c/J\approx2/[\ln(J/J') - \ln\ln(J/J')]$,
lead to a loss of monotonous convergence when $J'/J$ varies from
unity to zero.

We choose infinitely long clusters $L\times L\times\infty$
stretched in a lattice direction with the dominant interaction $J$.
Such a cluster geometry reflects the physical situation in the
system.
Therefore one may expect more precise results for the critical
temperature as the anisotropy of the quasi-one-dimensional
lattice increases.
We may also expect the monotonous convergence for the estimates
from Eq.~(\ref{eq:kappa2}) and (\ref{eq:chi4}) because there
are must be physical reasons (finite length of clusters
in the longitudinal direction, etc.)
for the non-monotonous or oscillatory character of behavior;
they are absent in our approximations.
That is, if Eq.~(\ref{eq:kappa2}) yields the lower bound in the
most unfavorable case $J'=J$ then it should preserve such
behavior for all $J'<J$.
Similar arguments are valid for the estimates following from
Eq.~(\ref{eq:chi4}); these estimates are upper.

Note that the mean values from Table~\ref{tab:Tc} are better not only
than the estimates of $k_BT_c/J$ calculated with the $(3,4)$ cluster
pair by the standard phenomenological RG method, but than their
improvements found by means of three-point extrapolations from
the sizes $L=2$, 3, and 4 to the bulk limit \cite{YuS91}.

In the range $10^{-2}\leq J'/J\leq1$, there are also the data
for the critical temperature of a simple-cubic Ising lattice which
were extracted from the Pad\'e-approximant analysis of the
high-temperature series \cite{NJ78}.
For $J'=J$ according to these data, $k_BT_c/J=4.5106$ that is lower
by $0.014\%$ in comparison with the results of Ref.~\cite{BST99}.
For $J'/J=0.1$ the authors of Ref.~\cite{NJ78} found the value
$k_BT_c/J=1.343$.
This quantity overestimates somewhat the mean value from
Table~\ref{tab:Tc}.
At last, for $J'/J=0.01$ the series method \cite{NJ78} yields
$k_BT_c/J=0.65$ that goes out of our lower bound.
This is not surprising because the calculations based on the
high-temperature series rapidly deteriorate owing to the very
limited number ($\le11$) of terms available in such series for the
anisotropic lattices.

So, we may treat the values found from Eqs.~(\ref{eq:kappa2})
and (\ref{eq:chi4}) as lower and upper bounds on the real
critical temperature.
Their mean value for each $J'/J$ yields the best estimate which
we achieve in this paper for the reduced critical temperature
$k_BT_c/J$ (the last column in Table~\ref{tab:Tc}).
By this, its absolute error is not larger in any case than
the half difference of the corresponding upper and lower bounds.
Using data from Table~\ref{tab:Tc} we establish that the relative
errors for $k_BT_c/J$ monotonically decrease from $0.72\%$ to
$0.14\%$ as $J'/J$ goes from 1 to $10^{-3}$.


\subsection{Invariants of the 3D Ising universality class}
\label{sec:I3DIUC}
Taking the improved estimates for the critical temperature of
anisotropic simple-cubic lattice we calculate now some invariants
of the three-dimensional Ising model universality class.


\subsubsection{Critical exponents}
\label{sec:EXPO}
According to the RG theory, critical exponents are determined entirely
by a fixed point and do not depend on the lattice anisotropy.
For a three-dimensional Ising model the universality of critical
exponents has been confirmed for $\Delta\in[0.2,5]$ by the
high-temperature series calculations \cite{PS72}.

At present, the most precise estimates of critical exponents
are provided by the high-temperature expansions for models with
improved potentials characterized by suppressed leading scaling
corrections.
For the fully isotropic 3D Ising lattice these methods yield
$\nu=0.630\,12(16)$ and $\gamma=1.2373(2)$ \cite{CPRV02}.
Hence, $y_t=1/\nu=1.5870(4)$ and $y_h=(d+\gamma/\nu)/2=2.481\,80(18)$.

In Table~\ref{tab:ytyh} we report our estimates for the critical
exponents $y_t$ and $y_h$.
It follows from those data that when the lattice anisotropy
parameter $\Delta$ varies in three orders (from unity to $10^{-3}$),
the estimates of critical exponents are changed only on a few per
cent or less.
In particular, calculations by Eqs.~(\ref{eq:yt1}) and (\ref{eq:yt2})
with the cluster pair $(3,4)$ yield respectively $y_t=1.47(6)$ and
$y_t=1.60(7)$.
Their variations are over the range $4-4.4\%$.
Similar calculations of the magnetic critical exponent 
carried out by use of Eqs.~(\ref{eq:yh1}) and (\ref{eq:yh2})
also with the pair $(3,4)$ leads to $y_h=2.586(5)$ and $y_h=2.579(5)$,
correspondingly.
Relative dispersions of these estimates are about $0.2\%$.

Thus, our calculations confirm the universality of both 
critical exponents in a remarkably wider range of $\Delta$ than it
was done in earlier investigations.
Systematic errors of the achieved estimates arise due to small sizes,
$L$, of subsystems used.


\subsubsection{Critical FSS amplitude ratios}
\label{sec:AMPL}
Critical amplitudes are determined by scaling functions.
As a result, their ``universal ratios'' like
$A_{\kappa^{(4)}}/A_{\chi^{(4)}}
={\cal K}^{(0,4)}(0,0)/{\cal F}^{(0,4)}(0,0)$ depend, generally
speaking, upon the lattice anisotropy because it can change the shape
of subsystems.
But in the case of parallelepipeds $L^{d-1}\times\infty$ with
unchanged (between themselves) transverse coupling constants the
shape of a sample (all its aspect ratios) will be independent of the
interaction in longitudinal direction.
Such a kind of the universality is studied here.

Table~\ref{tab:UY1Y2} contains results of our calculations for the
critical FSS amplitude ratios $U=A_{\chi^{(4)}}A_\kappa/A_\chi^2$,
$Y_1=A_{\kappa^{''}}/A_\chi$, and $Y_2=A_{\kappa^{(4)}}/A_{\chi^{(4)}}$.
Calculations have been carried out for $\Delta\in[10^{-3},1]$ by
use of a cyclic cluster $4\times4\times\infty$.

In accord with data of Table~\ref{tab:UY1Y2}, the average ratio
$U=4.900(3)$.
Hence, when the anisotropy parameter $\Delta$ varies in three orders,
this quantity changes only on $0.06\%$.
With such accuracy we may consider the given ratio as a constant.
In the case of fully isotropic lattice, $A_\kappa=1.26(5)$ and
$A_{\chi^{(4)}}/A_\chi^2=3.9(2)$ \cite{Yu97} and therefore
$A_{\chi^{(4)}}A_\kappa/A_\chi^2=4.9(5)$.
Our values of $U$ from Table~\ref{tab:UY1Y2} are in good agreement
with this estimate.

It is follow from Table~\ref{tab:UY1Y2} that
$Y_1=A_{\kappa^{''}}/A_\chi=1.759(2)$.
Hence, the constancy of this universal amplitude ratio is estimated at
least as a few times $10^{-3}$.
Our average value for $Y_1$ agrees well with the estimate for
the isotropic lattice, $A_{\kappa^{''}}/A_\chi=1.749(6)$ \cite{Yu97}.

According to data of Table~\ref{tab:UY1Y2} the amplitude ratio
$Y_2=A_{\kappa^{(4)}}/A_{\chi^{(4)}}=2.0133(6)$.
Thus, this quantity is most stable out of all invariants of the 3D
Ising universality class which were investigated in the paper.
Note that we are not aware of any quantitative estimates for the
$A_{\kappa^{(4)}}/A_{\chi^{(4)}}$.


\section{Conclusions}
\label{sec:Concl}
In this paper the large-scale transfer-matrix computations of the
phase-transition temperatures, as well as the critical exponents and
critical FSS amplitudes have been carried out.
Application of the extended phenomenological RG schemes allowed us
to find the tight limits on the critical temperature in the
anisotropic simple-cubic Ising lattice and improve the available
estimates for it.

We also calculated the thermal and magnetic critical exponents.
Our results confirm the universality of $y_t$ within $4-4.4\%$
and of $y_h$ within $0.2\%$ over a wide range of $\Delta$,
$10^{-3}\le\Delta\le 1$.

Finally, the presented results give an obvious evidence that the
critical FSS amplitude ratios
$U=A_{\chi^{(4)}}A_\kappa/A_\chi^2$, $Y_1=A_{\kappa^{''}}/A_\chi$,
and $Y_2=A_{\kappa^{(4)}}/A_{\chi^{(4)}}$ do not depend on the lattice
anisotropy parameter $\Delta=J'/J$ with accuracies at least $0.1\%$.
We give likely for the first time in the literature an estimate for
the universal quantity $Y_2$.


\section*{Acknowledgment}
This work was supported by the Russian Foundation for Basic Research
under project 03-02-16909.



\newpage
\begin{table}
\caption{
Lower and upper bounds on the critical temperature and their mean
values (our estimates of $k_BT_c/J$) in the 3D sc
spin-${1\over2}$ Ising lattice vs $\Delta=J'/J$.
Calculations with a cluster pair $(3,4)$.}
\label{tab:Tc}
\begin{center}
\begin{tabular}{lccc}
\hline\\[-2mm]
$\Delta$&${\rm Eq.}(4)$&${\rm Eq.}(5)$&mean\\[2mm]
\hline
  1.0    &4.47965814  &4.54424309  &4.51195062\\
  0.5    &2.91008665  &2.94295713  &2.92652189\\
  0.1    &1.33649605  &1.34570054  &1.34109829\\
  0.05   &1.03544938  &1.04144927  &1.03844933\\
  0.01   &0.65054054  &0.65323146  &0.65188600\\
  0.005  &0.55440490  &0.55643112  &0.55541801\\
  0.001  &0.40743000  &0.40859011  &0.40801006\\
\hline
\end{tabular}
\end{center}
\end{table}
\clearpage


\newpage
\begin{table}
\caption{
Estimates of the thermal and magnetic critical exponents by different
values of $J'/J$.
Calculations with a cluster pair $(3,4)$.}
\label{tab:ytyh}
\begin{center}
\begin{tabular}{lcccccc}
\hline\\[-2mm]
&&$\qquad\qquad\qquad y_t$&&&$\qquad\qquad\qquad y_h$&\\
&\\ \cline{3-4}\cline{6-7}\\
$J'/J$&$k_BT_c/J$&${\rm Eq.}(6)$&${\rm Eq.}(7)$&&${\rm Eq.}(8)$&
${\rm Eq.}(9)$\\[2mm]
\hline
   1.0   &4.51195062 &1.5760695 &1.7246286 &&2.5971647 &2.5886128\\
   0.5   &2.92652189 &1.5256373 &1.6636718 &&2.5902006 &2.5819462\\
   0.1   &1.34109829 &1.4700811 &1.5972576 &&2.5843305 &2.5766511\\
   0.05  &1.03844933 &1.4533899 &1.5791439 &&2.5836720 &2.5761101\\
   0.01  &0.65188600 &1.4236178 &1.5480583 &&2.5832982 &2.5758028\\
   0.005 &0.55541801 &1.4141719 &1.5383503 &&2.5832029 &2.5757888\\
   0.001 &0.40801006 &1.3984754 &1.5222765 &&2.5834573 &2.5757953\\[.5mm]
   &                 &1.47(6)   &1.60(7)   &&2.586(5)  &2.579(5)\\
\hline
\end{tabular}
\end{center}
\end{table}
\clearpage


\newpage
\begin{table}
\caption{Estimates of the universal critical FSS amplitude ratios
$U=A_{\chi^{(4)}}A_\kappa/A_\chi^2$,
$Y_1=A_{\kappa^{(2)}}/A_\chi$, and
$Y_2=A_{\kappa^{(4)}}/A_{\chi^{(4)}}$
for the Ising system with the cylindrical geometry
$L\times L\times\infty$ and periodic boundary conditions.
Data for $L=4$.}
\label{tab:UY1Y2}
\begin{center}
\begin{tabular}{lcccc}
\hline\\[-2mm]
$J'/J$ & $k_BT_c/J$ & $U$ & $Y_1$ & $Y_2$\\[2mm]
\hline
   1.0   &4.51195062   &4.8956599    &1.7550004   &2.0146443\\
   0.5   &2.92652189   &4.8967625    &1.7572512   &2.0136519\\
   0.1   &1.34109829   &4.9011909    &1.7596003   &2.0129829\\
   0.05  &1.03844933   &4.9014406    &1.7597697   &2.0129285\\
   0.01  &0.65188600   &4.9015375    &1.7598563   &2.0128977\\
   0.005 &0.55541801   &4.9015529    &1.7598646   &2.0128953\\
   0.001 &0.40801006   &4.9015782    &1.7598732   &2.0128938\\[.5mm]
   &                   &4.900(3)     &1.759(2)    &2.0133(6)\\
\hline
\end{tabular}
\end{center}
\end{table}


\end{document}